\documentclass[aps,prl,twocolumn,showpacs,amsmath,amssymb,10pt]{revtex4-1}
\usepackage{graphicx}
\usepackage{subfigure}
\usepackage{ulem}
\usepackage{amsmath}
\usepackage{amssymb}
\usepackage{wrapfig}
\usepackage[utf8]{inputenc}  %for accented characters in bibliography
\usepackage{color} %add color to a new command \fix

\begin{document}

\title{Atom-pair kinetics with strong electric-dipole interactions}
\author{N.~Thaicharoen$^{1\ast}$}
\author{L.F.~Gon\c{c}alves$^{1,2}$}
\author{G.~Raithel$^{1}$}
\affiliation{$^{1}$Department of Physics, University of Michigan, Ann Arbor, Michigan 48109, USA}
\affiliation{$^{2}$Instituto de F\'{\i}sica de S\~ao Carlos, Universidade de S\~ao Paulo, Caixa Postal 369, 13560-970, S\~ao Carlos, SP, Brasil}
\date{\today}

\begin{abstract}
Rydberg-atom ensembles are switched from a weakly- into a strongly-interacting regime via adiabatic transformation of the atoms from an approximately non-polar into a highly dipolar quantum state. The resultant electric dipole-dipole forces are probed using a device akin to a field ion microscope. Ion imaging and pair-correlation analysis reveal the kinetics of the interacting atoms. Dumbbell-shaped pair correlation images demonstrate the anisotropy of the binary dipolar force. The dipolar $C_3$ coefficient, derived from the time dependence of the images, agrees with the value calculated from the permanent electric-dipole moment of the atoms. The results indicate many-body dynamics akin to disorder-induced heating in strongly coupled particle systems.
\end{abstract}

%PACS numbers
%	32.80.Ee - Rydberg states
%	32.80.Rm - multiphoton excitation and ionization
%	34.20.Cf - interatomic potentials and forces
%32.80.Qk 	Coherent control of atomic interactions with photons
 %
%32.80.Rm 	Multiphoton ionization and excitation to highly excited states

\maketitle

%\section*{Introduction}
Dipolar and van der Waals interactions between atoms and molecules affect the properties of matter on microscopic and macroscopic scales. On the quantum level, the distinctions between van der Waals and electric dipole-dipole interactions are in the overall interaction strength, the scaling with the internuclear separation, and the (an)isotropy behavior. Highly excited Rydberg atoms present an ideal platform to study these interactions in binary and few-body quantum systems because Rydberg-atom interactions are generally strong and widely tunable between dipole-dipole, van der Waals and other types. The electric dipole-dipole~\citep{van_ditzhuijzen_spatially_2008, altiere_dipole-dipole_2011,ravets_measurement_2015} and van der Waals~\citep{reinhard_double-resonance_2008, beguin_direct_2013} interactions between Rydberg atoms have previously been studied using spectroscopic measurements of level shifts. Methods from optical and electron microscopy have been adapted to image Rydberg-atom systems with single-particle spatial resolution, revealing many-body quantum structures such as Rydberg-atom crystals~\citep{pohl_dynamical_2010, bijnen_adiabatic_2011, schaus_crystallization_2015} and enabling advanced studies of the Rydberg excitation blockade~\citep{lukin_dipole_2001, urban_observation_2009, weber_mesoscopic_2015}.

In our work we employ an adiabatic quantum-state preparation method and a modified field ion microscope~\citep{muller_field_1956, muller_field_1965, ullrich_recoil-ion_2003} with single-atom resolution to measure the atom kinetics that result from the dipolar force. We overcome the density limit imposed by the excitation blockade by initially preparing Rydberg atoms under conditions where they are only subject to weak van der Waals interactions. This allows us to prepare Rydberg atom samples with relatively small interatomic separations. In order to switch on strong dipole-dipole interactions, the atoms are subsequently transferred into a highly dipolar state via a Landau-Zener adiabatic passage through an avoided crossing~\citep{rubbmark_dynamical_1981, wang_dipolar_2015}. After the adiabatic state transformation, the direction of the permanent atomic dipoles is locked to the direction of an external electric field. After initialization, the atoms move under the influence of the strong dipolar forces for a variable interaction time. The center-of-mass positions of the Rydberg atoms are then detected by field ionization~\citep{gallagher_rydberg_2005} and ion imaging~\citep{schwarzkopf_imaging_2011}. From the recorded images, we calculate the spatial pair correlation functions between the Rydberg atoms~\citep{schwarzkopf_spatial_2013}, which directly show the anisotropic character of the dipolar force between two atoms and allow us to measure the  $C_3$ dipole-dipole interaction strength parameter. The experiment is performed under two conditions, one in which the direction of the atomic dipoles  is transverse and another in which it is perpendicular to the observation plane.

\begin{figure}[b]
\centering
\includegraphics[width=1\linewidth]{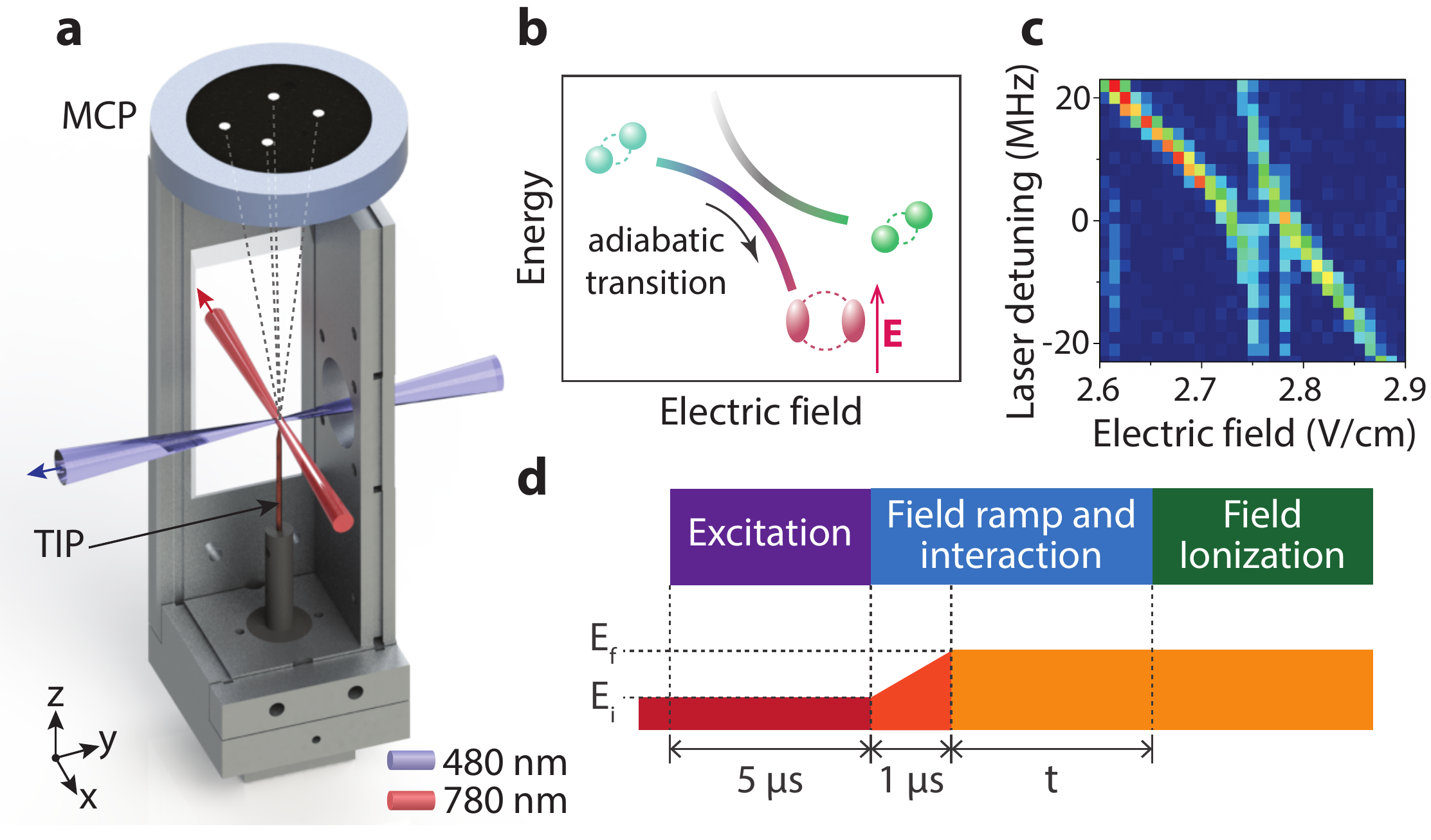}
\caption{(a) Experimental set-up. (b) Rydberg atoms initially prepared in an $S$-like state (blue circle) become adiabatically transferred  into a highly dipolar state (pink oval),  when passing through an avoided crossing in the Stark energy level  diagram in the applied electric field \bf E \rm. The diameters of the dashed circles indicate binary-interaction strengths. (c) Measured Stark spectra at the 6$^{\rm{th}}$ crossing between the $50S_{1/2}$ state and $n=47$ hydrogen-like states. (d) Timing sequence.}
\label{fig:figure1}
\end{figure}

\begin{figure*}
\centering
\includegraphics[width=0.7\textwidth]{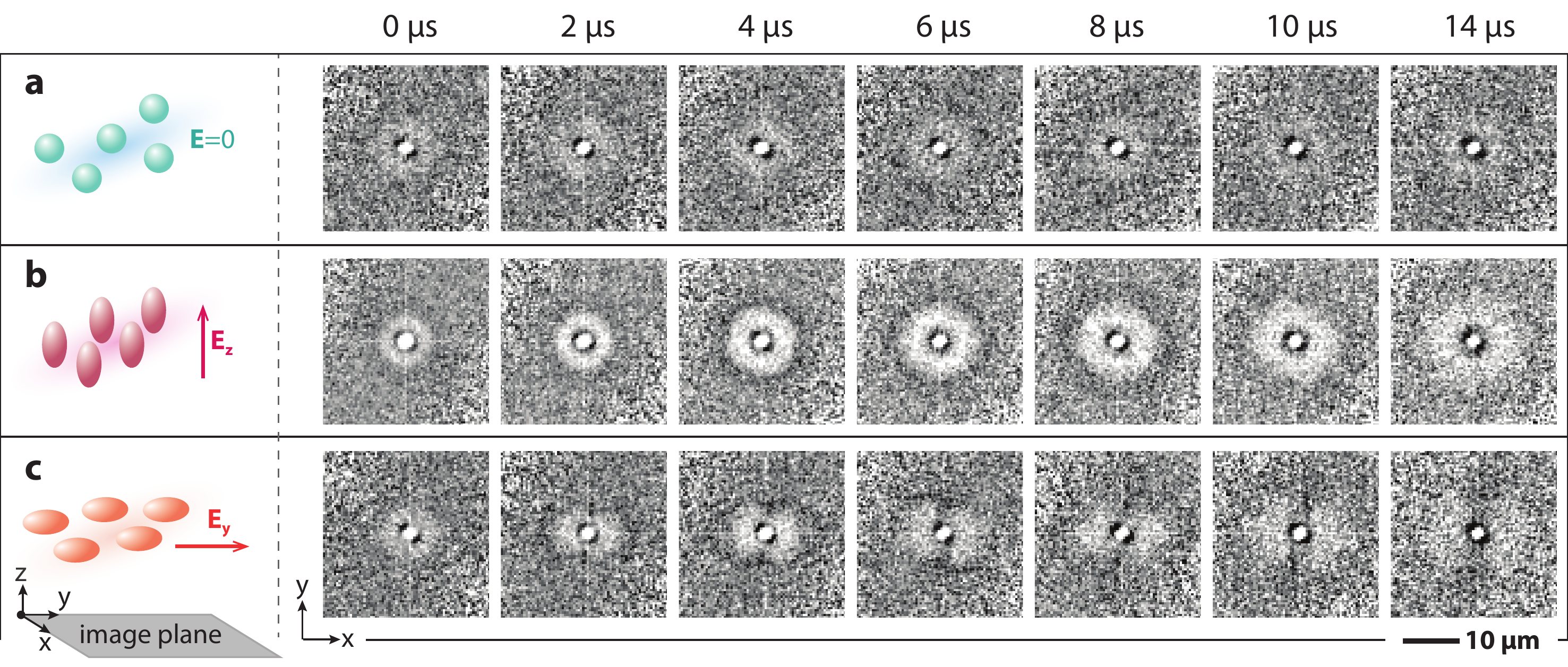}
\caption{Experimental pair correlation images for the indicated wait times. The linear grayscale ranges from 0.4 (white) to 1.5 (black), where values of 1, $<$1, and $>$1 indicate no correlation, anticorrelation, and positive correlation, respectively. 
The illustrations on the left show the electric-field and dipolar alignments relative to the object/image planes for the three data rows. (a) Zero field. Atoms undergo weak van der Waals interaction. (b) Applied electric field in $z$-direction. The atomic dipoles are aligned along $z$, perpendicular to the image ($xy$) plane. Binary interactions are azimuthally symmetric about $z$ and primarily repulsive.
(c) Applied electric field in $y$-direction. The atomic dipoles are aligned along $y$, in-plane with the image plane. Binary interactions are mixed attractive / repulsive, leading to anisotropic images showing repulsion along $x$ and attraction along $y$. The dark rings immediately around the centers are experimental artifacts.}
\label{fig:figure2}
\end{figure*}

%\section*{Experimental setup}
The experimental setup is shown in Fig.~\ref{fig:figure1}(a). Cold $^{85}$Rb atoms in the $5S_{1/2}$ state undergo two-photon Rydberg excitation into the $50S_{1/2}$ state by simultaneously applying 780- and 480-nm laser pulses with 5~$\mu$s duration and $\approx$1~GHz red-detuning from the 5$P_{3/2}$ intermediate state. The 780-nm beam propagates along the $x$-direction (coordinate frame defined in Fig.~\ref{fig:figure1}(a)) and has a Gaussian beam parameter of $w_{0}\approx 70~\mu$m. The 480-nm beam propagates in the $xy$ plane, forms an angle of approximately 70$^\circ$ with the 780~nm beam, and has $w_{0}\approx 8~\mu$m. Therefore, the excitation volume is cylindrical and aligned approximately along the $y$-direction. The number of Rydberg excitations in each sample is on the order of ten.

After excitation into the $50S_{1/2}$ state, the atoms are adiabatically transferred into a linear Stark state by sweeping the applied electric field through the avoided crossing between those two states~\citep{rubbmark_dynamical_1981, wang_dipolar_2015}. The permanent electric dipole moment of the Stark state is about 16 times larger than that of the initial (perturbed) $50S_{1/2}$ state, leading to much stronger, longer-range and anisotropic interactions, as sketched in Fig.~\ref{fig:figure1}(b). We perform the experiment at the 6$^{\rm{th}}$ avoided crossing (counted from zero applied field) of rubidium $50S_{1/2}$ with the manifold of $n=47$ hydrogen-like dipolar states, located at a calculated electric-field value of 2.762~V/cm. We use Stark spectroscopy to verify the position of this avoided crossing, as shown in Fig.~\ref{fig:figure1}(c). This particular avoided crossing is chosen because its gap size is large enough that we are able to experimentally resolve it in the Stark spectra, and it is large enough to ensure near-perfect adiabatic passage behavior.

The timing sequence of the experiment is shown in Fig.~\ref{fig:figure1}(d). The $50S_{1/2}$-like atoms, prepared at 0.1~V/cm below the avoided crossing, have a small electric dipole moment of $1.24\times10^{-27}$~Cm. They are adiabatically transferred into the dipolar target state by linearly increasing the electric field by 0.2~V/cm within 1~$\mu$s. The dipolar state has a dipole moment of $19.8\times10^{-27}$~Cm, corresponding to an increase of the dipolar interaction strength by a factor of 250.  The sweep duration is slow enough that the adiabatic transfer from the $50S_{1/2}$-like into the dipolar state has an efficiency very close to unity~\citep{method}. The sweep duration is also short enough that during the adiabatic transfer the atoms do not move significantly. Hence, the state switching is practically instantaneous with respect to the center-of-mass motion of the atoms. After the switch, the electric field is kept constant for the duration of the interaction time, $t$. We study the atom kinetics that follow from the dipolar force as a function of $t$.

To perform atom imaging, Rydberg atoms are ionized by sudden application of a positive high voltage to a tip imaging probe (TIP). Ions are accelerated by the TIP electric field towards a microchannel plate (MCP). Every detected ion results in a blip produced by the MCP-phosphor detector assembly, which reveals the center-of-mass position of its parent Rydberg atom at the time of ionization. The excitation region (object plane) is about 300~$\mu$m above the TIP. For this distance, the magnification is calibrated to be $200 \pm 10\%$~\citep{method}. We take 10 000 images in each data set. Using a peak-detection algorithm~\citep{schwarzkopf_spatial_2013}, the images are processed into a data structure in which each record contains the total number of detected ions and the ion impact positions in an individual image. The 5000 records with the highest ion numbers are processed into a sample-averaged pair correlation image, which is normalized such that at large distances it approaches the value of one. The average pair correlation images yield the information on the atom-pair kinetics.

%\section*{Electric field perpendicular to the detection plane}
For two identical dipoles pointing along the same direction, given by the direction of the applied electric field, the dipolar force has radial and polar components
\begin{equation}
\label{eq:eq2}
\begin{split}
F_{\rm{R}}&=\frac{{3p^{2}}}{R^{4}}[1-3\cos^{2}(\Theta)]\\
F_{\Theta}&=\frac{{-3p^{2}}}{R^{4}}[2\cos(\Theta)\sin(\Theta)]
\end{split}
\end{equation}
where $\Theta$ is the angle between the internuclear separation vector, $\bf{R}$, and the dipole vectors, $\bf{p}$.

When the applied electric field is perpendicular to the detection ($xy$) plane, the atomic dipoles point in the $z$-direction.
Since the excitation blockade radius is on the order of the diameter of the 480-nm excitation beam, for most dipole-dipole-interacting atom pairs the angle $\Theta$ in Eq.~\ref{eq:eq2} is $\approx\pi/2$. Hence, for the vast majority of atom pairs the dipole force is repulsive, with $F_{\rm{R}} \approx 3p^{2}/R^{4}$, and azimuthally symmetric about the line of sight. The pair correlation images at different interaction times, presented in Fig.~\ref{fig:figure2}(b), are azimuthally symmetric at all times. With increasing interaction time, the (projected) radius of the region that is largely devoid of pair-correlation events increases, reflecting an increase of the interatomic separation due to repulsive dipole-dipole interactions. We also find that the pair correlation is enhanced at a certain radius that increases in time; this radius approximately doubles over a time of 14~$\mu$s.

The expansion in Fig.~\ref{fig:figure2}(b), which is due to dipole-dipole interaction, is considerably faster than that due to repulsive van der Waals interactions between $50S_{1/2}$ atoms at zero electric field, shown in Fig.~\ref{fig:figure2}(a). The van der Waals interaction does not cause any significant expansion over the time scale in Fig.~\ref{fig:figure2}. In previous work~\citep{thaicharoen_measurement_2015} it was found that even $70S_{1/2}$ Rb Rydberg atoms, which interact about 50 times more strongly than $50S_{1/2}$ Rb Rydberg atoms (present case), exhibit significant repulsion effects only after about $30~\mu$s. Hence, a cursory comparison of the pair correlation data in Figs.~\ref{fig:figure2}(a) and~\ref{fig:figure2}(b) already shows that the interaction between the dipolar atoms must be one to two orders of magnitude stronger than the interaction between the non-polar, van der Waals-interacting atoms.

\begin{figure}[htb]
\centering
\includegraphics[width=\linewidth]{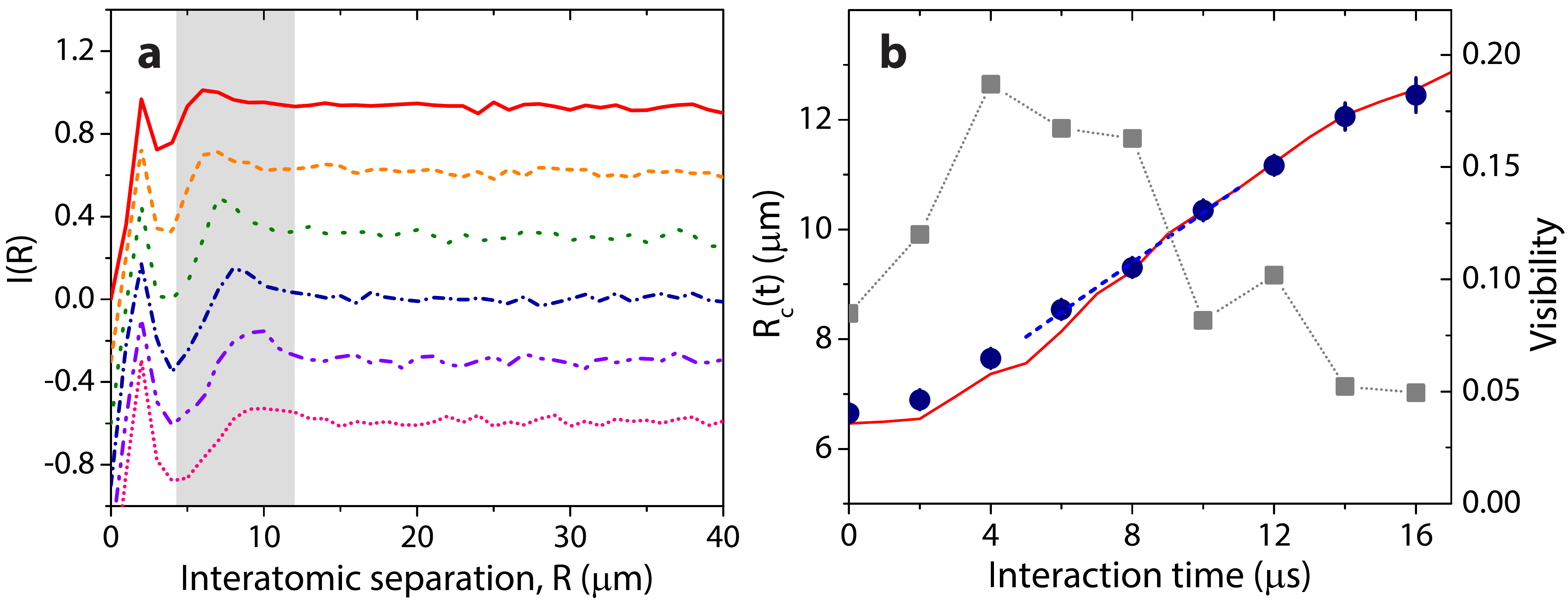}
\caption{(a) Angular integrals $I(R)$ of the pair correlation images in Fig.~\ref{fig:figure2}(b) at wait times (from top to bottom): 0~$\mu$s, 2~$\mu$s, 4~$\mu$s, 6~$\mu$s, 8~$\mu$s, and 10~$\mu$s. The $y$ axis is for the 0~$\mu$s curve; for clarity, the other curves are shifted down in equidistant steps of 0.3. (b) Interatomic separations $R_c(t)$ between Rydberg-atom pairs as a function of interaction time (left axis) from experiment (blue circles) and simulation (red line). The blue dashed line shows a linear fit to the experimental data between 6~$\mu$s to 10~$\mu$s. The gray squares show the visibility of the experimental pair correlation enhancement (right axis).}
\label{fig:figure3}
\end{figure}

The angular integrals $I(R)$ of the pair correlation images from Fig.~\ref{fig:figure2}(b) are shown in Fig.~\ref{fig:figure3}(a). The most probable separation between Rydberg-atom pairs, $R_c(t)$, is determined by local parabolic fits to $I(R)$ in the vicinity of the peaks found within the shaded region in Fig.~\ref{fig:figure3}(a). The blue circles in Fig.~\ref{fig:figure3}(b) show the resulting $R_c(t)$ values that represent the most probable radial atom-pair trajectory projected into the $xy$-plane. It is seen that atom pairs are initially prepared at a preferential separation $R_c(0)=6.7\pm0.7~\mu$m, controlled by the excitation-laser detuning and the atomic interaction strength before the adiabatic state transformation~\citep{schwarzkopf_spatial_2013}. The large positive acceleration observed subsequent to the state transformation is due to the strong repulsive dipole-dipole interaction that occurs for angles $\Theta$ near $\pi/2$. The acceleration apparently diminishes and changes sign from positive to negative at an interaction time near 8~$\mu$s, where $d^2 R_c(t) / dt^2 \sim 0$.

To verify that the observed rapid expansion in the $xy$-plane is consistent with the known permanent electric dipole moment of the atoms after adiabatic transformation, we extract the $C_{3}$ coefficient using conservation of energy  between interaction times  of $0$ and $t$,
\begin{equation}
\label{eq:eq3}
\frac{C_{3}}{R_c(0)^{3}}=\frac{C_{3}}{R_c(t)^{3}}+\frac{1}{2}\mu V(t)^{2} \quad .
\end{equation}
There, $\mu$ and $V(t)$ are the reduced mass of a pair of $^{85}$Rb atoms and the relative pair velocity, respectively. We choose $t=8~\mu$s because at that time the atom pairs have an approximately constant velocity, which can be extracted well from a local linear fit within the range $6~\mu$s $\le t \le $ 10~$\mu$s, indicated by the blue dashed line in Fig.~\ref{fig:figure3}(b). We obtain $V(8~\mu$s$)=0.45\pm0.06$~m/s and $R_c(8~\mu$s$)=9.3\pm0.9~\mu$m, where the uncertainties include the statistical fit uncertainty and the magnification uncertainty.  With $R_c(0)=6.7\pm0.7~\mu$m from above, the resulting $C_{3}$ value becomes $(3.3\pm1.8)\times10^{-42}~\rm{Jm}^{3}$. This value agrees with the calculated $C_{3}$ value, $p^2/(4 \pi \epsilon_0) = 3.55\times10^{-42}~\rm{Jm}^{3}$. This agreement also implies that the interacting entities are individual atoms and not superatoms, as has been predicted in~\citep{mobius_breakup_2013} and experimentally observed for van-der-Waals-interacting Rydberg excitations in~\citep{thaicharoen_measurement_2015}.

In Fig.~\ref{fig:figure3}(b) it is evident from the experimental data (circles) and the result of a semi-classical simulation~\citep{method} (red line) that the acceleration is negative for $t \gtrsim 10~\mu$s. The late-time deceleration appears to be due to repulsion from initially farther-away atoms, indicating many-body dynamics that involve more than two atoms. A related conclusion  can be drawn from considering the visibility of the pair-correlation enhancement as a function of time,
calculated from the $I(R)$ curves in Fig.~\ref{fig:figure3}(a) as $(I(R_{\rm{c}}(t))-\bar{I})/\bar{I}$, where the asymptotic values $\bar{I}(t)$ are obtained by averaging $I(R)$ curves over the range $R>15~\mu$m.
The visibility values are shown as gray squares in Fig.~\ref{fig:figure3}(b). The visibility rapidly increases at early times, $t \lesssim 4~\mu$s, and passes through a broad maximum between 4 and 8~$\mu$s. Hence, the significance of the pair correlation enhancement at $R_c$, equivalent to the degree of short-range order, and the kinetic energy in the sample, equivalent to the slope in $R_c(t)$, both become maximal approximately at the same time. These evidences are reminiscent of those in disorder-induced heating, which has been observed in the strongly-coupled ion component of an ultracold plasma~\citep{killian_ultracold_2007}. In both cases, particles initially repel each other due to dominant nearest-neighbor forces, before encountering repulsive forces from initially more distant particles. At wait times $t \gtrsim 10~\mu$s the correlation enhancement disappears, which is in part due to the initial thermal atom velocity (the magneto-optical trap has an atom temperature of $\sim 100~\mu$K). The ``coupling parameter'' of the dipolar system can be defined as the ratio between the initial dipole-dipole interaction energy and the thermal energy at $100~\mu$K. This ratio is $\Gamma = 9$, which is sufficiently large that the system should indeed develop (transient) short-range order.

%\section*{Electric field parallel to the detection plane}
When the electric field is applied along $y$, the atomic permanent electric dipole moments are oriented along $y$. In that case, the angular dependence of the dipole force can be observed within the $xy$ plane, encompassing maximally-repulsive interactions ($\Theta = \pi/2$) and maximally-attractive interactions ($\Theta = 0$), as well as all intermediate cases. The interaction then leads to the characteristic anisotropic patterns in the pair correlation images shown in Fig.~\ref{fig:figure2}(c).

\begin{figure}[htb]
\centering
\includegraphics[width=\linewidth]{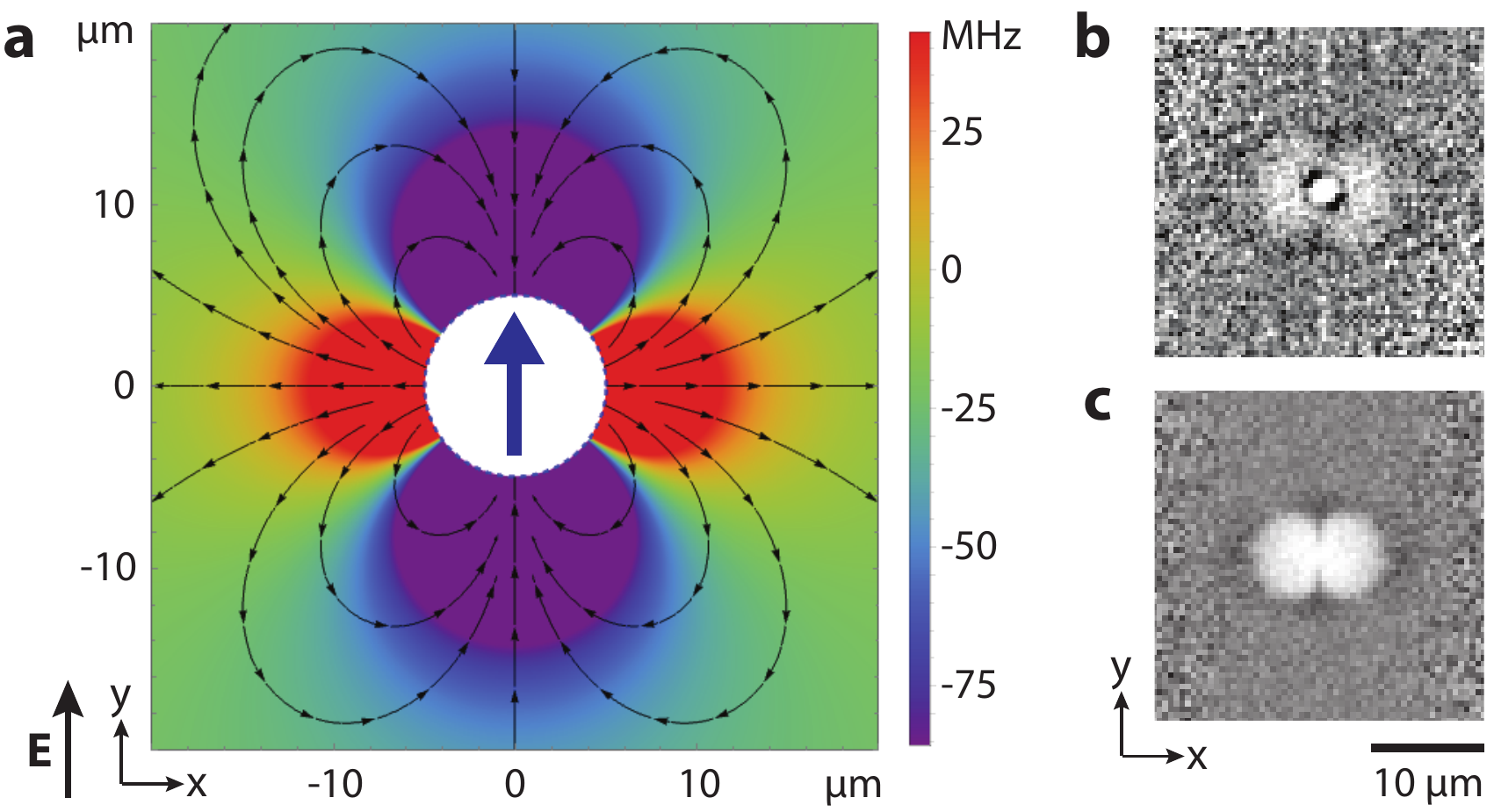}
\caption{(a) Color map of the dipolar potential and corresponding force vectors as a function of $\bf{R}$ for a pair of dipoles pointing along $y$. Atoms repel each other in the equatorial direction, $x$, and attract each other in the polar direction, $y$. Due to the combination of radial and angular forces, the atom pairs become funneled into  narrow conical sections at the poles. This leads to characteristic pair correlation images as shown in (b) (from the experiment) and (c) (from a simulation). The images in (b) and (c) are for $t=4~\mu$s.}
\label{fig:figure4}
\end{figure}

The angular force in Eq.~\ref{eq:eq2} is maximal at $\Theta=\pi/4$ and $3 \pi/4$, zero at $\Theta=0$, $\pi/2$ and $\pi$, and it always points towards the ``poles'' (see Fig.~\ref{fig:figure4}(a)). Hence,  the angular force leads to an accumulation of atom pairs lined up close to the electric-field direction (the same direction as $\bf{p}$). These atom pairs then become pulled close to each other due to the radial component of the force, which is attractive for $\Theta \le 55^\circ$ and $\Theta \ge 125^\circ$. These atom pairs form the prominent  vertical dark strip across the center of the images at interaction times $\gtrsim 4~\mu$s in Fig.~\ref{fig:figure2}(c). Conversely, atom pairs initially positioned at $\Theta \approx \pi/2$ will keep repelling each other, while being diverted towards the poles by the angular force (see Fig.~\ref{fig:figure4}(a)). Within the experimental uncertainty, the most probable pair separations $R_c(t)$ in Fig.~\ref{fig:figure2}(b), along any direction in the $xy$ plane, and in Fig.~\ref{fig:figure2}(c), along the $x$ direction, are the same. This is expected because all these cases correspond to $\Theta=\pi/2$ in Eq.~\ref{eq:eq2}. The ``funneling effect'' pointing towards the poles eventually leads to dumbbell-shaped pair correlation images that are void of signal in a volume extending along $x$ and that possess enhanced signals along $y$. These characteristics are seen in the experiment (Fig.~\ref{fig:figure2}(c) and Fig.~\ref{fig:figure4}(b)) and in our simulations (Fig.~\ref{fig:figure4}(c)). The small deviation of the enhancement cones in Fig.~\ref{fig:figure2}(c) from the $y$-direction is attributed to a slight deviation of the electrode arrangement from perfect symmetry.

Atoms pulled close to each other along the polar direction ($y$) will likely undergo Penning-ionizing collisions~\citep{robicheaux_ionization_2005}, when the interatomic separation $R$ drops below about 0.5~$\mu$m in our case. This distance is below the image resolution and is not directly observed in the experiment. It is, however, noticed that the amount of signal within the enhancement cones near $\Theta=0$ and $\Theta=\pi$ in Fig.~\ref{fig:figure2}(c) plateaus at $t \gtrsim 4~\mu$s and eventually drops. This observation is consistent with atom-pair loss within the polar cones due to Penning ionization.

%\section*{Conclusion}
In summary, we have employed an adiabatic state transformation method to prepare dense samples of about ten  Rydberg atoms with large permanent electric dipole moments. We have determined the  dipolar dispersion coefficient, $C_3$. The measured pair correlation images portray the anisotropic character of dipolar atom-pair kinetics in a forceful, intuitive manner.  Results of a semi-classical trajectory model agree well with our experimental observations. We have observed dynamics reminiscent of disorder-induced heating, as found elsewhere in strongly-coupled plasmas. Further insight may be gained by varying the ratio between the time scales governing the atoms' center-of-mass motion and the dipolarization (the ramp speed used during adiabatic passage). This may lead into a new approach for the formation of ordered quantum matter, such as crystalline states of Rydberg atoms.

\begin{acknowledgments}
This work was supported by NSF Grant No. 1205559 and 1506093, and AFOSR Grant No. FA9550-10-1-0453.  N.T. acknowledges support from DPST of Thailand. L.F.G. acknowledges support from FAPESP Grant No. 2014/09369-0.
\end{acknowledgments}

\bibstyle{apsrev4-1}
%\bibliography{dipoleref}


\end{document}

\baselineskip16pt

\maketitle
\section*{}

\noindent \bf Experimental setup\\[6pt]
\indent\rm See Fig. 1 in the main text for an illustration. $^{85}$Rb atoms are prepared in a magneto-optical trap at a density of $\sim 10^{10}$~cm$^{-3}$ and a temperature of $\sim 100~\mu$K. Approximately 15 atoms are excited into the $50S_{1/2}$ Rydberg state by using coincident 780-nm and 480-nm laser pulses. Both beams provide an excitation volume that is small enough to lessen the effect of electric-field inhomogeneity near the tip imaging probe (TIP) but large enough to contain about ten $50S_{1/2}$ Rydberg atoms. The Rydberg-atom number is limited by the weak van der Waals excitation blockade that applies to the $50S_{1/2}$ state. The electric field within the excitation volume is aligned along the $y$- or $z$-direction (see Fig.~1(a)). During the laser excitation, the initial field strength $E_{\rm{i}}$ is set 0.1~V/cm below $E_{\rm{c}}$, the field at which the 6$^{\rm {th}}$ avoided crossing between the $50S$-like Stark state and the hydrogenic manifold is centerd ($E_c = 2.762$~V/cm). After laser excitation, the electric field is ramped linearly to a final value of $E_{\rm{f}} = E_{\rm{c}} + 0.1$~V/cm.  The ramp time is 1~$\mu$s. The field is held at $E_{\rm{f}}$ while we wait for the atom sample to evolve over the selected times. At the end of the interaction time, we apply a sudden 1.6~kV potential to the TIP, which is a beryllium-copper needle with a rounded head of 125~$\mu$m diameter. The Rydberg atoms become field-ionized, and the resultant ions are imaged onto a microchannel plate detector (detection efficiency 30$\%$ - 50$\%$, corresponding to an average number of about 5 detected Rydberg atoms). The ion impact positions generate blips on the phosphor screen of the detector assembly, providing a spatial image of the ion (and atom) distribution. The images are recorded and analyzed as detailed below.\\

\noindent \bf Adiabatic transfer through an avoided crossing\\[6pt]
\indent\rm Here it is not possible to directly excite highly dipolar, elongated Stark states at a well-defined initial separation $R_c(0)$ that is sufficiently small to allow for kinetic studies of dipolar interactions. This is due to the Rydberg excitation blockade, which limits the achievable density, and the low oscillator strength of linear Stark states for optical excitation.
To solve this problem, we initially laser-excite atoms into an $S$-like Rydberg state in a weak electric field. This state has an oscillator strength that is sufficiently large for optical excitation of about ten Rydberg atoms during a time of $5~\mu$s. Also, since atoms in the $S$-like Rydberg state mostly interact via a weak isotropic van der Waals potential, the Rydberg-atom density limit imposed by the excitation blockade is quite high ($\sim 10^9~$cm$^{-3}$). We adiabatically transfer the $S$-like Rydberg atoms into a highly dipolar, linear Stark state by the means of an adiabatic passage through a level crossing~\cite{wang_dipolar_2015}. Here we are working at the 6$^{\rm{th}}$ crossing between the $50S_{1/2}$ Stark state and the hydrogenic manifold. This crossing has a gap size of 42~MHz, which is large enough that we can resolve the crossing using Stark spectroscopy. The probability of adiabatic state transformation is obtained from the Landau-Zener formula, which leads to (in SI units)
\[P_{\rm adia} = 1-\exp(-\Delta t / \tau) \quad {\rm{with}} \quad \tau = \frac{\hbar \Delta E \vert p_f - p_i \vert}{2\pi(\Delta W/2)^2} \quad . \]
There, $\Delta E$ is the range of the linear electric-field sweep (0.2~V/cm in our case), $\vert p_f - p_i \vert$ is the difference in electric dipole moment of the intersecting states (here $p_i=1.24\times10^{-27}$~Cm, $p_f=19.8 \times 10^{-27}$~Cm), $\Delta W$  is the minimal gap splitting at the center of the avoided crossing (here $h \times 42$~MHz), and $\Delta t$ is the ramp duration (here 1~$\mu$s). It follows $\tau = 32$~ns and $P_{\rm adia} = 1-2.5\times 10^{-14}$, so that the probability of adiabatic transfer is close to unity. The adiabatic state transformation results in an enhancement of dipolar interaction strength by a factor of about 250.\\

\noindent \bf Electric field calibration\\[6pt]
\indent\rm To control the electric field inside the chamber, we apply voltages to a set of electrodes and the TIP. We perform Stark spectroscopy of the $48D_{J}$ levels in order to zero and calibrate the electric field in the excitation region. The field zero corresponds to the symmetry lines of the Stark energy level diagrams measured as a function of the applied voltages. The applied field is zeroed and calibrated along all three axes. The voltage-to-electric-field calibration factors are obtained by comparing measured and calculated spectra. A convenient field marker is the field at which the $\vert nD_{5/2}, m_{j}=\pm 1/2 \rangle$ levels become stationary as a function of applied field. For $n=48$, this occurs at an electric field of 2.08~V/cm. \\

\noindent \bf Magnification calibration\\[6pt]
\indent\rm We temporarily insert a double-slit mask in the collimated 480-nm beam. The mask is placed in front of the lens that focuses the beam into the experimental chamber. The slits have a width of 1.0~mm and a center-to-center separation of $d=6.0$~mm. The separation between adjacent diffraction maxima for wavelength $\lambda=480$~nm and focal length of the lens $D=20$~cm is $\lambda D/d=16~\mu$m. This value has been verified by images of the double-slit diffraction pattern, taken at the focal point of the lens. The distribution of Rydberg-atom counts obtained with the double-slit mask in place consists of a series of parallel, equidistant stripes, which correspond to the maxima of the diffracted blue excitation light. To obtain a smooth image of the Rydberg-atom distribution, we average over at least $2 \times 10^3$ individual images. The average spacing between the parallel stripes in the averaged image is 80 pixels, leading to a calibration factor of 5.0$\pm$0.5~pixels per $\mu$m distance in the object (atom) plane. This scaling factor is used to generate the length markers in the figures. Since one pixel of the camera corresponds to a distance of 40.5~$\mu$m on the detector assembly's phosphor screen, the TIP magnification is 200, with a relative uncertainty of $10\%$.\\

\noindent \bf Spatial pair correlation function and angular integration\\[6pt]
\indent\rm For each individual image, we apply a peak-detection algorithm to obtain the ion positions in that image. Each individual image yields a pixelated pair correlation image $X_{i,j}^{(k)}$, where $i,j$ are pixel indices, $k$ is an image counter, and the value of $X_{i,j}^{(k)}$ is the number of occurrences that the separation vector between two ions resides within the pixel $(i,j)$. Note the images $X_{i,j}^{(k)}$ are only very sparsely populated. We then apply an area correction factor that corrects for the fact that, for a given frame size $N\times M$, pixels $(i,j)$ with larger magnitudes of $i$ and $j$ have smaller image overlaps when executing the pair correlation. The area-corrected pair correlation images are $A^{(k)}_{i,j}=f(i,j,N,M) X_{i,j}^{(k)}$, where $f(i,j,N,M)=NM/[(N-\abs{i})(M-\abs{j})]$ is the area correction factor. Then we calculate the average pair correlation function $A=(1/K) \sum\limits_{k=1}^{K} A^{(k)}$, where $K$ is the number of images used in the average. Moderate coarse-graining over an area approximately equivalent to the point spread function of a single-ion blip yields the pair correlation images $\bar{A}$ shown in Fig.~2. Azimuthal integration of the $\bar{A}$ yields the radial pair correlation functions $I(R)$ in Fig.~3.\\

\noindent \bf Semi-classical calculation\\[6pt]
\indent\rm We perform a semi-classical 3D simulation of the dynamics of Ryd\-berg atoms interacting via dipole-dipole forces.  Rydberg-atom positions and velocities are initialized as explained in~\cite{thaicharoen_measurement_2015}, and the positions and velocities are then propagated using a Runge-Kutta integrator that includes all pair-wise interatomic forces. For the initialization of the simulated samples, we use $C_{6}=1.014\times10^{-59}$~Jm$^{6}$ (calculated in~\cite{reinhard_level_2007}), and in the Runge-Kutta integrator we use $C_{3}=3.546\times10^{-42}~\rm{Jm}^{3}$ (obtained from the electric-dipole moments evident in the Stark spectra). In the simulations we record the atom positions at selected interaction times. We then calculate the average pair-correlation images and radial functions $I(R)$ using the same procedure as in the experiment.\\

\bibliography{dipoleref}
\bibliographystyle{Science}

\clearpage